\documentclass[%
reprint,
superscriptaddress,
nofootinbib,
amsmath,amssymb,
aps,
pra,
]{revtex4-2}

\usepackage{hyperref}
\usepackage{graphicx}% Include figure files
\usepackage{dcolumn}% Align table columns on decimal point
\usepackage{bm}% bold math
\usepackage{placeins}
\usepackage{enumitem}
\usepackage{multirow}
\usepackage{orcidlink}

\newcommand{\xiph}{$\xi_{\mathrm{ph}}(r) $}

\begin{document}

\preprint{APS/123-QED}
\title[Protohalos and scale-dependent bias for BAO]{Leveraging protohalos and scale-dependent bias\\to calibrate the BAO scale in real space}

\author{Sasha Gaines \orcidlink{0000-0002-2240-7421}}
\email{sasha.gaines@yale.edu}
\affiliation{Department of Astronomy, Yale University, New Haven, CT 06511, USA}

\author{Farnik Nikakhtar \orcidlink{0000-0002-3641-4366}}
\affiliation{Department of Physics, Yale University, New Haven, CT 06511, USA}

\author{Nikhil Padmanabhan \orcidlink{0000-0002-2885-8602}}
\affiliation{Department of Physics, Yale University, New Haven, CT 06511, USA} 
\affiliation{Department of Astronomy, Yale University, New Haven, CT 06511, USA}

\author{Ravi K.~Sheth \orcidlink{0000-0002-2330-0917}}
\affiliation{Center for Particle Cosmology, University of Pennsylvania, Philadelphia, PA 19104, USA}
\affiliation{The Abdus Salam International Center for Theoretical Physics, Strada Costiera 11, Trieste 34151, Italy}

\date{\today}

\begin{abstract}
The location of the baryon acoustic oscillation (BAO) feature in the two-point correlation function (2PCF) of matter produces a standard ruler that is useful for the measurement of the expansion history of the Universe. Inspired by the possibility of reconstructing the positions of protohalos in the initial density field with a novel method rooted in optimal transport theory, we revisit the BAO signal in the protohalo correlation function.
Our work examines the performance of a template 2PCF built on a tracer bias relation that includes scale dependence -- a term that can be motivated by peaks theory or a general bias expansion. Working in protohalos, halos, and the linear combination of the protohalo and matter fields that is motivated by the continuity equation, we demonstrate that this model accurately captures the shape of the BAO feature and improves the precision of the BAO scale measurement relative to a model that does not include scale-dependent bias by 47\% in protohalos, 15\% in halos, and 14\% in the linear combination of the protohalo and matter fields. Allowing for scale dependence does not appear to introduce any shift in the BAO feature. The precision of the BAO distance scale estimate is highest with the linear combination of the protohalo and matter fields, which offers a factor of 3.5 improvement over Eulerian-space measurements and a factor of 4-8 improvement over the estimate made with protohalos alone.
\end{abstract}

\maketitle

{\centering
\begin{quotation}
A ring inscribes eternal marvels\\
    The first, the name of which is dawn,\\
    Sends worlds awhirl about an axis\\
    Borne of reality's self-governed flow.

    -- Veronika Tushnova \footnote{V. Tushnova. \textit{Krug vechnyī tainstv dvukh idët.} Interpreted for this epigraph by the first author.}

\end{quotation}
}
\section{Introduction}

The clustering of matter holds an imprint of the acoustic waves that rolled through the primordial fluid until photon-baryon decoupling \cite{Eisenstein1998}. In real space, that imprint takes the form of a peak in the two-point correlation function of matter. This peak, termed the configuration-space BAO feature, is linked to the sound horizon at recombination. This deterministic nature gives the BAO scale promise as a standard ruler, useful for measuring the Hubble parameter, angular diameter distance, and the expansion history of the Universe \cite[e.g.][]{Cooray2001, Amendola2005, Blake2003, Hu2003, Linder2003, Seo2003, Matsubara2004, Blake2005, Cole2005, Eisenstein2005, Glazebrook2005, Blake2006, Zhan2006, Padmanabhan2008, Martinez2009, Shoji2009}. Since we cannot observe the unbiased distribution of matter directly, we turn to the correlation function of its biased tracers to measure the BAO scale in data. 

While halos are a generally accepted stepping stone between unbiased matter and biased galaxies \cite{Cooray2002a}, working in Eulerian space presents the need to treat the effects of nonlinear evolution, which smear and shift the BAO signature \cite{Eisenstein2007, Crocce2008}. The standard approach relies on the reconstruction of the initial linear matter field based on the observed positions of galaxies \cite{Eisenstein2007}. While this technique has largely focused on improving the accuracy of BAO measurements \cite[e.g.][]{Padmanabhan2012a, Paillas2024}, it has not emphasized reconstructing the initial positions of halos. A recent variant of reconstruction, based on optimal transport theory, explicitly aims to do just that, and has demonstrated that it is possible to reconstruct initial halo positions accurately \cite{Nikakhtar2022}. This work explores the possibility of doing a BAO analysis with these reconstructed positions by studying the actual ``protohalo'' positions -- the centers-of-mass of regions in the initial Gaussian random field of matter that are destined to form halos. Additionally, we put forth a multi-tracer approach to BAO analyses by combining protohalos with the matter overdensity field. This new dataset aims to amplify the benefits offered by protohalos while boosting signal-to-noise.

Defining a model that characterizes the shape of the expected correlation function is a basic step in the measurement of the BAO scale. The linear theory two-point correlation function of the dark matter field is the fundamental building block of this model. However, the shape of the evolved correlation function is significantly different from linear theory, especially around the acoustic peak region \cite{Crocce2008}.  Moreover, for biased tracers, even the initial shape differs from that of the dark matter \cite{Desjacques2008, Desjacques2010, Desjacques2010a, Musso2012, Baldauf2017}. This difference motivates a model correlation function that accounts for shape variations, in addition to a simple change in amplitude.

To this end, we pursue a model correlation function with a bias operator that adds a $k^2$ term to the scale-independent linear bias factor. The fact that the scale-dependent bias factor multiplies $k^2$ (rather than another power of $k$) can be motivated in two ways. First, $k^2$ arises as the first nonzero term (after the scale-independent $b_{10}$) from general bias expansions in $k$ that are constrained only by symmetries \cite[e.g. rotational and Galilean invariance, see Ref.][]{Desjacques2018}.  In such approaches, $k^2$ arises from a Laplacian, so is sometimes termed `derivative' bias.  On the other hand, in the simplest physically motivated `peaks'-based models of bias \cite[e.g.][]{Bardeen1986}, this form is exact -- there are no other terms of higher order in $k$.  In these models, the $k^2$ behavior arises from the constraint that the curvature of the overdensity of matter be negative at locations where halos are destined to form. In other words, the Laplacian has an explicit physical origin, and there are no other contributions to the bias if there are no other constraints. This $k^2$ scaling remains in the more recent and elaborate `excursion set peaks' halo formation frameworks \citep{Paranjape2012, Paranjape2013, Castorina2016, Nikakhtar2018, Musso2021}. 

We examine such a scale-dependent bias term's impact on the quality of fit and robustness of a BAO scale measurement in protohalos, halos, and a novel dataset that combines protohalos and the matter overdensity field. Section \ref{sec:theory} establishes the theoretical background. In Section \ref{sec:fit results}, we fit the model correlation function to simulation data and compare the results with a model that does not include scale-dependent bias, as well as the bias values we expect from the peak formalism. Section \ref{sec: conclusions} discusses our findings.  Two Appendices provide useful details, and a third, Appendix~\ref{appx:r0}, shows the impact of scale-dependent bias on the zero-crossing scale of the two-point correlation function.  

\begin{table*}
    \caption{Summary of the models of $b_1(k)$ (Eq.~\ref{eq:basic overdensity relation}) used in this paper. $b_{10}$ and $b_{01}$ refer to Lagrangian bias in all expressions.}
    \label{tab:b1}
    \centering
    
    \begin{tabular}{c @{\hskip 0.2in} c @{\hskip 0.2in} c}
 \hline \hline
    \\[-1em]
         &  \textbf{Full model}& \textbf{Simplified model} \\ \hline
    \\[-1em] \textbf{Protohalos} &  $(b_{10}+b_{01}R_v^2k^2)\Tilde{W}(kR_f)$& $b_{10}\Tilde{W}(kR_f)$ \\ 
    \\[-1em] \textbf{Protohalos$+\mathbf{\delta_{\rm{m}}}$} & $\left(b_{10} + b_{01} R_v^2 k^2\right)\,\Tilde{W}(kR_f) + 1$
    & $b_{10} \Tilde{W}(kR_f)+1 $ \\
    \\[-1em] \textbf{Halos} & $\left[b_{10} +1 + \left(b_{01} -1\right)R_v^2 k^2\right]\Tilde{W}(kR_{\rm smear}) $  & $(b_{10}+1)\Tilde{W}(kR_{\rm smear})$ \\ 
\hline \hline
    \end{tabular}
\end{table*}

\section{Theory and background}\label{sec:theory}
We begin with the ansatz that $\delta_{\rm{t}}$, the number density contrast of a biased tracer (such as a halo or its progenitor, a protohalo), can be predicted from the overdensity of matter, $\delta_{\rm{m}}$, via a bias operator. Assuming a linear bias operator, we can write the overdensity relation
\begin{equation}\label{eq:basic overdensity relation}
    \delta_{\rm{t}}(\vec{k}) = b_1(k) \delta_{\rm{m}}(\vec{k}),
\end{equation}
which we use to construct that tracer's model two-point correlation function (hereafter, 2PCF)\footnote{Note the abbreviated notation for isotropic spherical integration over all $\vec{k}$ values:
$ \int \frac{d^3 k}{(2 \pi)^3} \rightarrow \int_0^{\infty} \frac{k^2}{2\pi^2} dk \rightarrow \int_{\boldsymbol{\mathrm{k}}}.$}
\begin{equation}\label{eqn:xi general}
\xi_{\textrm{t}}(r)= \int_{\boldsymbol{\mathrm{k}}} b_{1}(k)^2 
\,P_{\textrm{Lin}}(k)\, j_0(k r),
\end{equation}
where $P_{\textrm{Lin}}(k)$ is the linear theory power spectrum at redshift $z^{\rm{E}}$, which corresponds to the epoch at which we identify halos ($z^{\rm{E}}= 0.1$ in this work); $j_0$ is the zeroth-order spherical Bessel function of the first kind.

Table \ref{tab:b1} summarizes the bias models in this work: those with a scale-dependent bias term (`Full model', Sec.~\ref{sec:scale-dep bias}) and models that only include a constant scalar bias term (`Simplified model', Sec.~\ref{sec:fitting without b01}).

\subsection{Scale-dependent bias}\label{sec:scale-dep bias}

\subsubsection{Protohalos}

We will consider bias models in which, for protohalos,
\begin{equation}\label{eqn: ph overdensity}
\delta_{\textrm{ph}}(\vec{k})=\left(b_{10} + b_{01} R_v^2 k^2 \right) \Tilde{W}(k R_f) \delta_{\textrm{m}}(\vec{k})
\end{equation}
\cite[e.g.][]{Desjacques2010a}.
Following common practice, we refer to $b_{10}$ as the scale-independent, linear bias factor. Nonzero $b_{01}$ leads to $k$-dependence often termed `scale-dependent' bias. Throughout this work, $b_{10}$ and $b_{01}$ denote Lagrangian bias.
$\Tilde{W}$ is the Fourier-space window function, for which we assume the Gaussian form
\begin{equation}\label{eqn:window}
    \Tilde{W}(kR_f)=\exp \left[-\frac{(k R_f)^2}{2}\right],
\end{equation}
although recent work suggests that $\Tilde{W}(x)=15 j_2(x)/x^2$ might be more physically motivated \citep{Musso2021}. The filter scale $R_f$ roughly corresponds to the radius of a patch of the early Universe (with average matter density $\bar{\rho}$), which would collapse into a halo of mass $M$ at a later time. For the Gaussian window, this sets 
\begin{equation}\label{eqn:Rf scale}
  R_f =\frac {1}{\sqrt{2 \pi}} \Bigg(\frac{M}{\bar{\rho}} \Bigg)^{1/3}.
\end{equation}
The particular choice of a Gaussian and the exact definition of $R_f$ is somewhat arbitrary; \cite{Paranjape2013} explores other reasonable choices. Although in symmetry expansions $R_v$ is a free parameter, in peaks-based models it has a clear physical meaning:  it is a characteristic scale that is related to the typical mass of the objects in the sample.  It has the form 
\begin{equation}
    R_v = \frac{\sigma_0(R_f)}{\sigma_1(R_f)},
\end{equation}
with $\sigma_0$ and $\sigma_1$, the spectral moments of the linear matter power spectrum smoothed on scale $R_f$:
\begin{equation}\label{eqn: spectral moments}
 \sigma_n^2(R_f) \equiv \int_{\boldsymbol{\mathrm{k}}} k^{2 n} \tilde{W}^2(k R_f) P_{\textrm{Lin}}(k).
\end{equation}

\begin{figure}
\centering
\includegraphics[width=0.97\columnwidth]{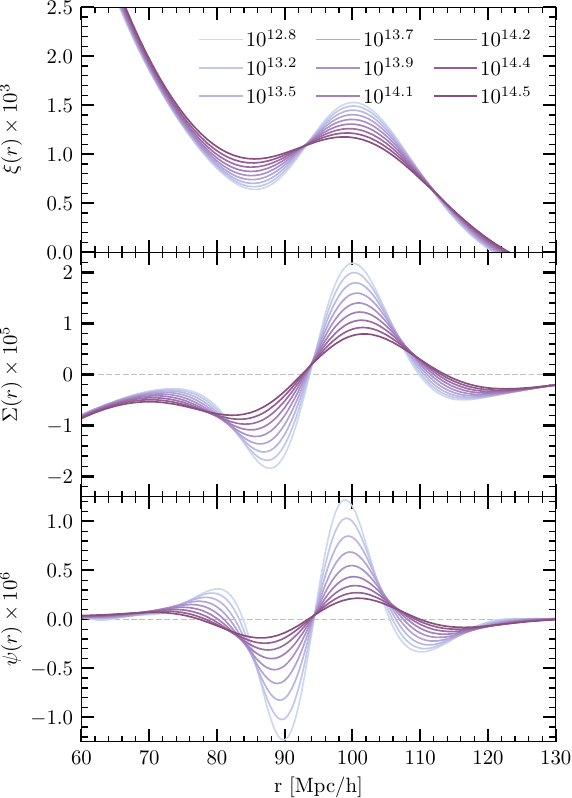}
\caption{Components of \xiph: Eq.~\ref{eqn:xi_subcomponent} (top panel),  Eq.~\ref{eqn:Sigma_subcomponent} (middle panel), and  Eq.~\ref{eqn:psi_subcomponent} (bottom panel). Colors indicate the masses, in $M_\odot /h$, which correspond to the $R_f$ scales used to compute each curve.
}
\label{fig:subcomponent}
\end{figure}

\begin{figure}
\centering
\includegraphics[width=0.97\columnwidth]{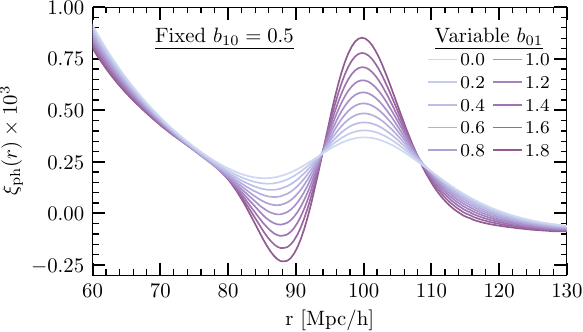}
\caption{ Illustration of the effect of varying $b_{01}$ on the shape of the correlation function. $R_f$ is set to 2.2 Mpc/$h$, $b_{10}=0.5$.}
\label{fig: bias illustration}
\end{figure}

Eq.~\ref{eqn:xi general} and Eq.~\ref{eqn: ph overdensity} give the protohalo correlation function
\begin{equation}\label{eqn:xi model}
\xi_{\textrm{ph}}(r)= \int_{\boldsymbol{\mathrm{k}}} \left(b_{10}+b_{01} R_v^2 k^2 \right)^2 \Tilde{W}^2(k R_f) 
\,P_{\textrm{Lin}}(k)\, j_0(k r)  .
\end{equation}

To foreshadow the results in Section~\ref{sec:fit results}, we break \xiph \ into three components:
\begin{align}\label{eqn:xi_subcomponent}
\xi(r)&= \int_{\boldsymbol{\mathrm{k}}}P_{\textrm{Lin}}(k)  \Tilde{W}^2(kR_f)j_0(k r), \\
\label{eqn:Sigma_subcomponent}
\Sigma(r)&=\int_{\boldsymbol{\mathrm{k}}} k^2 P_{\textrm{Lin}}(k) \Tilde{W}^2(kR_f) j_0(k r),\ \rm{and} \\
\label{eqn:psi_subcomponent}
\psi(r)&= \int_{\boldsymbol{\mathrm{k}}} k^4 P_{\textrm{Lin}}(k) \Tilde{W}^2(kR_f) j_0(k r),
\end{align}
which reduce to Eq.~\ref{eqn: spectral moments} at $r=0$: $\xi(0)=\sigma^2_0(R_f)$, $\Sigma(0)=\sigma^2_1(R_f)$ and $\psi(0)=\sigma^2_2(R_f)$. These components let us isolate the parts of Eq.~\ref{eqn:xi model} that $b_{01}$ and $b_{10}$ impact:
\begin{equation}\label{eqn:xi_in_components}
\xi_{\textrm{ph}}(r)=b_{10}^2 \xi(r) + 2 b_{10} b_{01}R_v ^ 2 \Sigma(r) + b_{01}^2R_v^4 \psi(r).
\end{equation}
Fig.~\ref{fig:subcomponent} shows each component's contribution to the total shape of \xiph \ as a function of mass. Smaller masses create more pronounced peaks in all three functions. Since $\tilde{W}$ is a smoothing of the density field in configuration space on a scale $R_f$, it isn't surprising that the prominence of the BAO feature decreases with increasing mass.

Even though the amplitudes of $\Sigma(r)$ and $\psi(r)$ are two-to-three orders of magnitude smaller than $\xi(r)$, factors of $R_v^2$ and $R_v^4$ ensure that they exert a non-negligible impact on \xiph. Fig.~\ref{fig: bias illustration} illustrates this: with $b_{10}$ held fixed, increasing $b_{01}$ sharpens the BAO peak in \xiph. In Sec.~\ref{sec:fit results}, we will see the consequences of including these $b_{01}$-dependent elements in the model we fit to data.

\subsubsection{Halos}
For halo bias, we add a smearing term by using $R_{\rm smear}$ as the smoothing scale in the Gaussian window function:
\begin{equation}\label{eqn:Rsmear}
 R_{\rm smear}^2 = \frac{2}{3}\,\sigma_{-1}^2(R_f)\,\Bigl[1 - \gamma_v^2(R_f)\Bigr] ,
\end{equation}
with $\gamma_v\equiv \sigma_0^2/(\sigma_{-1}\sigma_1)$ \cite{Bardeen1986}. %BBKS eq. 4.26
The smearing arises because of the displacements from Lagrangian to Eulerian positions \cite{Bharadwaj1996a, Crocce2008}, and the term $\sigma_{-1}^2$ is the 3D-velocity dispersion of matter particles in the Zel'dovich approximation. The factor of 1/3 makes this a 1D dispersion, and the factor of 2 accounts for the fact that we are working with pairs rather than single particles. The $\gamma_v$ correction comes from peaks theory; it arises because velocities and gradients of the density field are correlated \cite{Bardeen1986, Desjacques2010, peakVels}. 

Additionally, following \cite{Desjacques2010a}, we relate the Eulerian bias factors to their Lagrangian counterparts: 
\begin{subequations}
\begin{eqnarray}
           b_{01}^E &=& b_{01} - 1  \qquad {\rm and}\qquad \label{eqn:b01E}  \\
b_{10}^E &=& b_{10} + 1. \label{eqn:b10E}
    \end{eqnarray}
\end{subequations}
We thus relate $\delta_{\rm{m}}$ to the number overdensity of halos:
\begin{equation}\label{eqn:halo overdensity}
\delta_{\textrm{h}}(\vec{k})=\left\{\left[b_{10} +1 + \left(b_{01} -1\right) R_v^2 k^2\right]\, 
\Tilde{W}(kR_{\rm smear})\right\}\, \delta_{\textrm{m}}(\vec{k}) .
\end{equation}
Curly brackets here enclose the halo $b_1(k)$, which we combine with Eq.~\ref{eqn:xi general} to obtain the halo 2PCF, $\xi_{\rm{h}}(r)$.

Strictly speaking, Eq.~\ref{eqn:halo overdensity} should include a $\Tilde{W}(kR_f)$ term, which would suppress power at large $k$ -- but there should also be a mode-coupling piece, which would add power at large $k$. We set $W(kR_f)\to 1$ as an approximate way of including the mode-coupling-like contribution. Since $R_{\textup{smear}}^2\gg R_f^2$, the presence or lack of an $R_f$ term in Eq.~\ref{eqn:halo overdensity} impacts our fits negligibly. Our model also leaves out $b_2$, with the assumption that with $\delta \ll 1$ on BAO scales, $b_2$ makes a small contribution to halo bias, relative to the terms in Eq.~\ref{eqn:halo overdensity}. However, a rigorous examination of this assumption is left up to future work. 

\subsubsection{Combined protohalo and matter overdensity field}

In the context of Fig.~\ref{fig: bias illustration}, Eq.~\ref{eqn:b01E} gives rise to the possibility that the BAO feature is sharper in protohalo data than in halos -- a potential advantage for the measurement of the acoustic scale. Eq.~\ref{eqn:b10E}, however, reveals that Lagrangian bias is close to 0 when Eulerian bias is close to 1. This decreases the signal-to-noise in the 2PCF of datasets built on protohalos alone, as we will see in Sec.~\ref{sec:full model fit}. If additional steps are not taken, low signal-to-noise may diminish the advantage of an enhanced peak.

There may be a way to recover the amplitude of the signal while harnessing protohalos' sharper peak: \cite{Nikakhtar2024} recently showed that adding an estimate of the matter density field to the protohalo field can create an enhanced BAO feature. We thus add protohalo data to the linear matter overdensity field, scaled to the Eulerian redshift (see Sec.~\ref{sec:simulation} for the practical details). To model this combined field's 2PCF, we start with the relation
\begin{equation}\label{eqn: combined overdensity}
\begin{split}
\delta&_{\textrm{comb}}(\vec{k}) = \delta_{\textrm{ph}}(\vec{k})+\delta_{\textrm{m}}(\vec{k}) =\\
 &\left[\left(b_{10} + b_{01} k^2R_v^2\right)\,\Tilde{W}(kR_f)\, + 1\right]\, \delta_{\textrm{m}}(\vec{k}),
\end{split}
\end{equation}
where square brackets enclose the $b_1(k)$ term that we combine with Eq.~\ref{eqn:xi general} to obtain $\xi_{\rm{comb}}(r)$. In the context of Eq.~\ref{eqn:b10E}, the addition of 1 to the combined field's $b_1(k)$ leads us to expect the clustering of this combined field to match the amplitude of the halo 2PCF. Sec.~\ref{sec:full model fit}  will demonstrate that this does indeed happen in data.

While protohalos present a physically motivated choice for $R_f$ (c.f. Eq. \ref{eqn:Rf scale}), the choice of filter scale is not as straightforward for the combined field. We tested the fitting procedure (described in Sec.~\ref{sec:fit results}) on the combined field data with seven filter scale values, from $0.6R_f$ to $1.3R_f$. A scale equal to $R_f$ produced the lowest $\chi^2$, supporting the assumption that $R_f$ is the Lagrangian radius of a protohalo and is the appropriate filter scale for a model of the protohalo + $\delta_{\rm{m}}$ field.

Our analysis uses simulation data, which provides the `ground truth' $\delta_{\rm{m}}$ from the initial conditions (IC). Real observations, however, cannot access a ground truth. Instead, an observational BAO analysis would need to work with a reconstructed matter field, which would include noise and smoothing that the simulation IC lack. To get an idea of the impact of smoothing on BAO measurements, we will additionally apply the formalism in Eq.~\ref{eqn: combined overdensity} to the density field smoothed on scale $R_{\delta}$,
\begin{equation}\label{eq: smoothed delta}
    \delta_{\rm{m}}^W(\vec{k}) = \delta_{\rm{m}}(\vec{k}) \exp \left( \frac{-k^2 R_{\delta}^2}{2} \right).
\end{equation}
We leave a more detailed modeling of the estimated  $\delta_{\rm{m}}$ to future work.

\section{Fitting a bias model to data}\label{sec:fit results}
\begin{table}
\caption{Scales and spectral moments associated with the mass bins in this work. Length is in units of Mpc$/h$.}
\label{tab:mass bin properties}
 \centering
\resizebox{\columnwidth}{!}{
\begin{tabular}{ c c c c c c c}
\hline \hline \\[-1em]
Mass [$M_\odot/h$] & 
$R_f$  &
$R_v$ &
$R_{\rm{smear}}$ &
$\sigma_0$ &
$\sigma_1$ &
$\sigma_{-1}$ \\ \hline
\\[-1em] $10^{13.0}$-$10^{13.2}$ &
    1.886 & 
    2.499 &
    6.994 &
    1.187 &
    0.475 &
    9.096\\ $10^{13.2}$-$10^{13.4}$ &
    2.199 & 
    2.855 &
    6.776 &
    1.086 &
    0.380 &
    9.006\\
    $10^{13.4}$-$10^{13.6}$ &
    2.564 & 
    3.261 &
    6.496 &
    0.989 &
    0.303 &
    8.902\\
     $10^{13.6}$-$10^{13.8}$ &
    2.989 & 
    3.724 &
    6.129 &
    0.898 &
    0.241 &
    8.783\\
     $10^{13.8}$-$10^{14.0}$ &
    3.485 & 
    4.251 &
    5.639 &
    0.812 &
    0.191 &
    8.650  \\ \hline \hline
    
\end{tabular}}
\end{table}
 
\subsection{Simulation data}\label{sec:simulation}
We perform our fits on 25 (2 Gpc$/h)^3$ boxes of the {\sc{AbacusSummit}} suite of simulations \cite{Garrison2021, Maksimova2021} at the fiducial flat $\Lambda$CDM cosmology, with $\left(\Omega_m, \Omega_b, h\right)=(0.3152, 0.0507, 0.6736)$ and $2.1 \times 10^9 \ \textup{M}_\odot/h$ particles.

For halo positions, we use the centers of mass of particles that belong to halos at $z^{\rm{E}}=0.1$ in the {\texttt{compaSO}} catalogs \cite{Hadzhiyska2021}. We match these particles' IDs to their locations in the IC at $z^{\rm{IC}}=99$. The barycenters of these Lagrangian locations become the protohalo positions. 

To produce the combined protohalo+$\delta_{\rm{m}}$ field, we
\begin{enumerate}[label=(\arabic*)]
    \item scale the {\sc{AbacusSummit}} IC density field on a $1152^3$-cell mesh to $z^E$ by the growth factor $D(z^{\rm{E}})/D(z^{\rm{IC}})$;
    \begin{enumerate}
     \item optionally, smooth $\delta_{\rm{m}}$ to get $\delta_{\rm m}^W$ (Eq.~\ref{eq: smoothed delta});
    \end{enumerate}
    \item convert protohalo positions into a number overdensity mesh using {\texttt{nbodykit}}\footnote{\url{https://github.com/bccp/nbodykit}}\cite{Hand2018};
    \item add the meshes to get the combined overdensity:
    $\delta_{\rm{comb}} = \delta_{\rm{m}} + \delta_{\rm{ph}}$.
\end{enumerate}

We measure 2PCFs over 2 Mpc/$h$ radial bins from 39.5 to 139.5 Mpc/$h$ with {\texttt{nbodykit}}. Table \ref{tab:mass bin properties} lists our $R$ scales and spectral moments (Eq.~\ref{eqn: spectral moments}). $R_f$ is computed using each bin's median mass. The other values are computed using that $R_f$.  Appendix \ref{sec:covariance matrix} describes our covariance matrix and lists halo number densities.

\subsection{Standard BAO template}

To measure the BAO scale in data, we need a template function for the 2PCF. This template must:
\begin{enumerate}[label=(\arabic*)]
    \item account for our theoretical model of the physics behind the formation of the BAO feature;
    \item include effects that impact the shape of the 2PCF;
    \item not correlate with effects that physically influence the location of the acoustic peak.
\end{enumerate}
In an approach that has been used in past observational BAO analyses \citep[e.g.][]{Taylor2010, Ahn2012, Bautista2020, Philcox2021, Philcox2022}, we combine  Eq.~\ref{eqn:xi general} with $a_n$, nuisance parameters designed to capture influences on the 2PCF's shape that are not intrinsically relevant to extracting the acoustic scale. In this work, $n$ runs from 0 to 2. We marginalize over the nuisance parameters analytically (Appendix \ref{sec:nuisance parameters}). We get the template
\begin{equation}\label{eqn:standard template}
    \xi_{\rm{template}}(\alpha r)=\xi_{\rm{t}}(\alpha r) + \sum_n{\frac{a_n}{r^n}},
\end{equation}
where $\alpha$ acts as a proxy for the BAO scale, $r_*$:
\begin{equation}\label{eqn:alpha def}
\alpha\equiv\frac{D_{V}(z) r_{*}^{\mathrm{fid}}}{D_{V}^{\mathrm{fid}}(z) r_{*}},
\end{equation}
where `fid' indicates reference values from a fiducial cosmology, for which $\alpha$ equals 1. An $\alpha$ of 0.9 shifts the BAO peak 10\% higher than $r_{*}^{\mathrm{fid}}$; an $\alpha$ of 1.1 has the opposite effect. The spherically averaged distance $D_V$ is the distance that isotropic BAO measurements are sensitive to. When measured on data, $\alpha$ probes the background expansion history. In simulations (like this work) where the cosmology is known, $\alpha$ measures just how standard the BAO feature is. Eq.~\ref{eqn:standard template} is the currently standard functional form for measuring the BAO scale in large-scale surveys such as eBOSS \citep{Bautista2020}. In Sec.~\ref{sec:full model fit}, we fit Eq.~\ref{eqn:standard template} to data using $\xi_{\rm t}$ from the full model developed in Sec.~\ref{sec:scale-dep bias}. Sec.~\ref{sec:fitting without b01} repeats the fit, using a simplified model that does not include scale-dependent bias.

\begin{figure}
    \centering
    \includegraphics[width=\columnwidth]{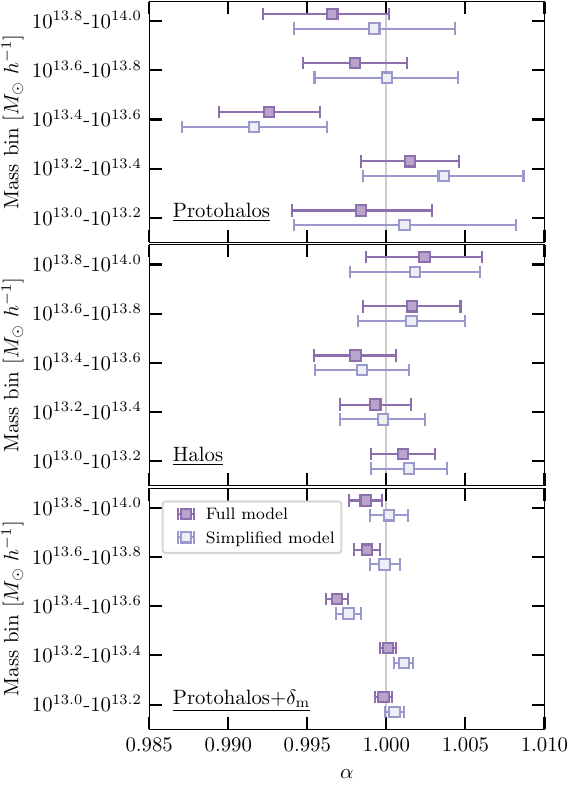}
    \caption{Best fit of $\alpha$, measured with protohalos (top panel), halos (middle panel) and the protohalo + $\delta_{\rm m}$ field (bottom panel). Error bars show $1\sigma$ from the MCMC chain's median.}
    \label{fig:alpha best fit}
\end{figure}

\begin{figure*}
    \centering
\includegraphics[width=0.98\textwidth]{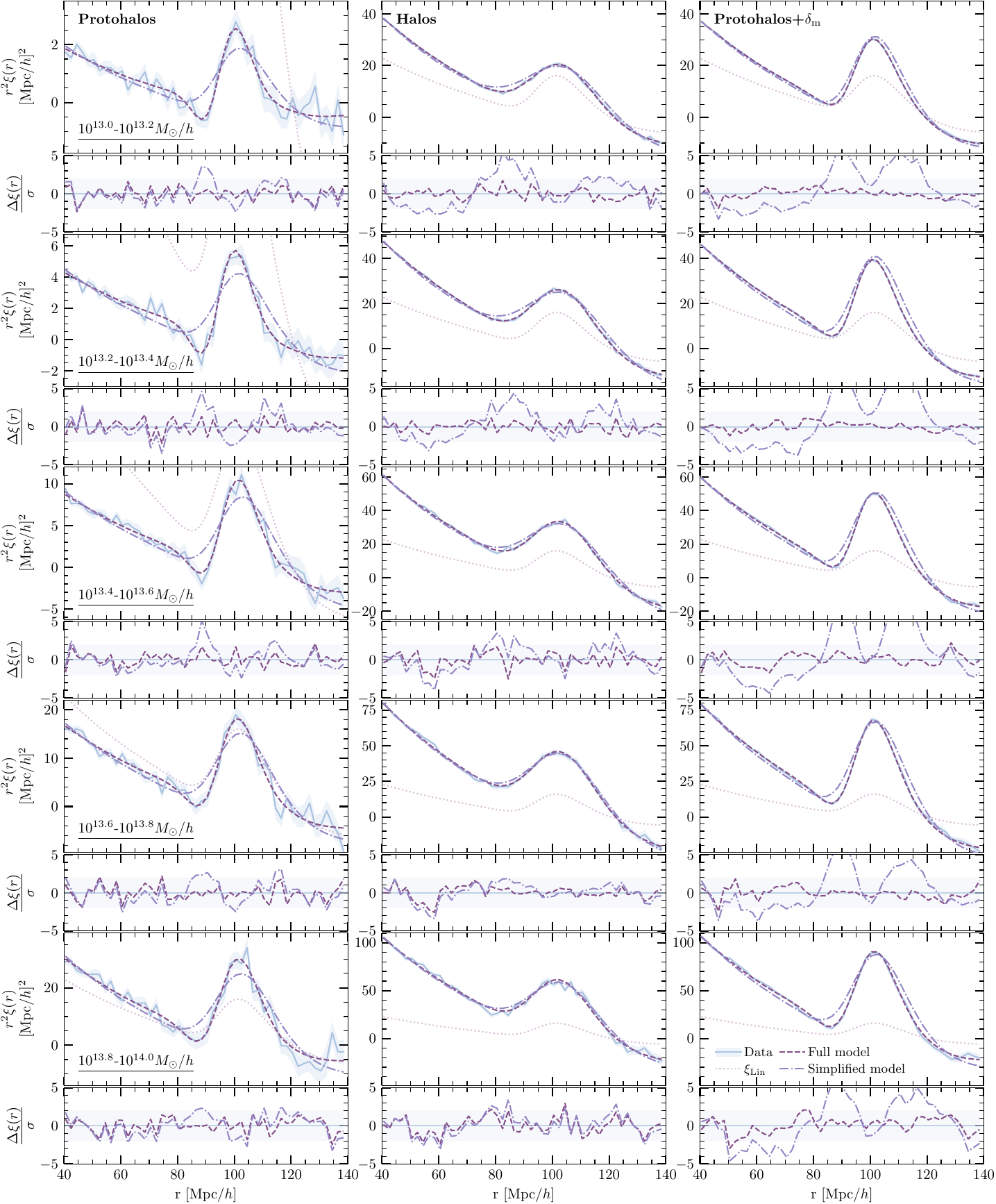}
\caption{\textbf{Odd rows}: Best-fit 2PCF for the full model (dashed line) and the simplified model (dot-dashed line) with best-fit parameters measured with protohalos (left panel), halos (middle panel) and the combined protohalo + $\delta_{\rm{m}}$ field (right panel), compared to the linear theory $\xi_{\rm{Lin}}$ (dotted line, same in all panels) and the mean {\sc{AbacusSummit}} $\xi \pm 1\sigma$ (solid line and light blue band around it). Mass bins are indicated in each row. \textit{Note}: the `Halos' and `Protohalos + $\delta_{\rm{m}}$' columns have the same vertical axis range. \textbf{Even rows}: residuals normalized by standard deviation $\sigma$, with the $2\sigma$ range highlighted in blue.}
\label{fig:merged results}
\end{figure*}

\begin{figure*}
\centering
\includegraphics[width=\textwidth]{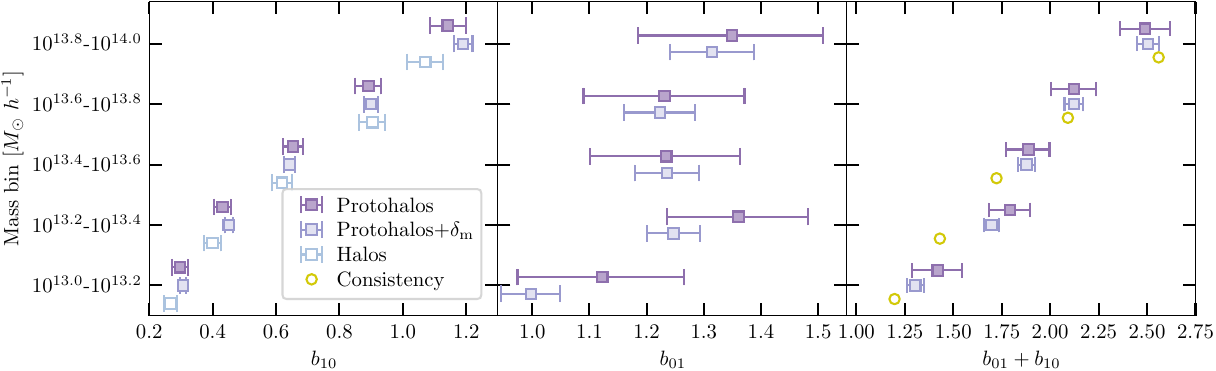}
\caption{Best fit for $b_{10}$ (left panel), $b_{01}$ (middle panel) and $b_{10}+b_{01}$ (right panel), measured with protohalos (dark squares), combined protohalo + $\delta_{\rm{m}}$ field (light squares) and halos (empty squares), as well as the sum predicted by theory in Eq.~\ref{eqn:consistency relation} (circles). Eulerian-space measurements of $b_{01}$ are omitted because they fall outside the reasonably expected range.}
\label{fig:bias values summary}
\end{figure*}

\subsection{Fitting the full bias model}\label{sec:full model fit}
\begin{table}
    \caption{Fit results. Best-fit values are the MCMC chain medians; errors are the 16th and 84th percentiles.}
    \label{tab:merged fit results}
    \resizebox{\columnwidth}{!}{
    \begin{tabular}{c c c @{\hskip 0.09in} c@{\hskip 0.09in} c @{\hskip 0.09in}c }
        \hline \hline \\[-1em]
        Tracer & Mass [$M_\odot/h$] & $\chi^2$ &  $b_{01}$ & $b_{10}$ & $\alpha$ 
        
      \\[-1em]  \\ \hline \hline     \\[-1em]
    \multicolumn{6}{|c|}{Full model}\\ \hline \hline
         \\[-1em] \multirow{6}{*}{\begin{tabular}[c]{@{}c@{}}Protohalos\end{tabular}} & $10^{13.0}$-$10^{13.2}$ & 1215 & $1.12_{-0.15}^{+0.14}$ & $0.30_{-0.02}^{+0.02}$ & $0.9984_{-0.0044}^{+0.0045}$ \\[-1em]  \\ \cline{2-6}
    \\[-1em]  & $10^{13.2}$-$10^{13.4}$ & 1206 & $1.36_{-0.13}^{+0.12}$ & $0.43_{-0.03}^{+0.03}$ & $1.0015_{-0.0031}^{+0.0031}$ \\[-1em]  \\ \cline{2-6}
    \\[-1em]  & $10^{13.4}$-$10^{13.6}$ & 1205 & $1.23_{-0.13}^{+0.13}$ & $0.65_{-0.03}^{+0.03}$ & $0.9926_{-0.0032}^{+0.0032}$ \\[-1em]  \\ \cline{2-6}
    \\[-1em]  & $10^{13.6}$-$10^{13.8}$ & 1228 & $1.23_{-0.14}^{+0.14}$ & $0.89_{-0.04}^{+0.04}$ & $0.9980_{-0.0033}^{+0.0033}$ \\[-1em]  \\ \cline{2-6}
    \\[-1em]  & $10^{13.8}$-$10^{14.0}$ & 1230 & $1.35_{-0.16}^{+0.16}$ & $1.14_{-0.06}^{+0.06}$ & $0.9966_{-0.0044}^{+0.0036}$ \\ \\[-1em] \hline \hline
    \\[-1em] \multirow{6}{*}{\begin{tabular}[c]{@{}c@{}}Halos\end{tabular}} & $10^{13.0}$-$10^{13.2}$ & 1232 & $4.51_{-0.29}^{+0.28}$ & $0.27_{-0.02}^{+0.02}$ & $1.0011_{-0.0020}^{+0.0020}$ \\[-1em]  \\ \cline{2-6}
    \\[-1em]  & $10^{13.2}$-$10^{13.4}$ & 1207 & $4.03_{-0.28}^{+0.27}$ & $0.40_{-0.03}^{+0.03}$ & $0.9993_{-0.0022}^{+0.0022}$ \\[-1em]  \\ \cline{2-6}
    \\[-1em]  & $10^{13.4}$-$10^{13.6}$ & 1237 & $3.01_{-0.26}^{+0.26}$ & $0.62_{-0.03}^{+0.03}$ & $0.9980_{-0.0026}^{+0.0026}$ \\[-1em]  \\ \cline{2-6}
    \\[-1em]  & $10^{13.6}$-$10^{13.8}$ & 1220 & $2.19_{-0.26}^{+0.26}$ & $0.90_{-0.04}^{+0.04}$ & $1.0016_{-0.0031}^{+0.0031}$ \\[-1em]  \\ \cline{2-6}
    \\[-1em]  & $10^{13.8}$-$10^{14.0}$ & 1232 & $2.15_{-0.26}^{+0.25}$ & $1.07_{-0.06}^{+0.06}$ & $1.0024_{-0.0037}^{+0.0037}$ \\ \\[-1em] \hline \hline
    \\[-1em] \multirow{6}{*}{\begin{tabular}[c]{@{}c@{}}Protohalos\\+ $\delta_{\rm{m}}$\end{tabular}} & $10^{13.0}$-$10^{13.2}$ & 1275 & $1.00_{-0.05}^{+0.05}$ & $0.31_{-0.01}^{+0.01}$ & $0.9998_{-0.0005}^{+0.0005}$ \\[-1em]  \\ \cline{2-6}
    \\[-1em]  & $10^{13.2}$-$10^{13.4}$ & 1231 & $1.25_{-0.05}^{+0.05}$ & $0.45_{-0.01}^{+0.01}$ & $1.0001_{-0.0005}^{+0.0005}$ \\[-1em]  \\ \cline{2-6}
    \\[-1em]  & $10^{13.4}$-$10^{13.6}$ & 1259 & $1.24_{-0.06}^{+0.06}$ & $0.64_{-0.02}^{+0.02}$ & $0.9969_{-0.0007}^{+0.0007}$ \\[-1em]  \\ \cline{2-6}
    \\[-1em]  & $10^{13.6}$-$10^{13.8}$ & 1243 & $1.22_{-0.06}^{+0.06}$ & $0.90_{-0.02}^{+0.02}$ & $0.9988_{-0.0008}^{+0.0008}$ \\[-1em]  \\ \cline{2-6}
    \\[-1em]  & $10^{13.8}$-$10^{14.0}$ & 1267 & $1.31_{-0.07}^{+0.07}$ & $1.19_{-0.03}^{+0.03}$ & $0.9987_{-0.0010}^{+0.0010}$ \\ \\[-1em] \hline \hline
    \\[-1em] \multirow{6}{*}{\begin{tabular}[c]{@{}c@{}}Protohalos\\+ $\delta_{\rm{m}}^W$\end{tabular}} & $10^{13.0}$-$10^{13.2}$ & 1298 & $0.90_{-0.06}^{+0.06}$ & $0.31_{-0.01}^{+0.01}$ & $0.9996_{-0.0007}^{+0.0007}$ \\[-1em]  \\ \cline{2-6}
    \\[-1em]  & $10^{13.2}$-$10^{13.4}$ & 1255 & $1.16_{-0.06}^{+0.06}$ & $0.45_{-0.01}^{+0.01}$ & $0.9999_{-0.0007}^{+0.0007}$ \\[-1em]  \\ \cline{2-6}
    \\[-1em]  & $10^{13.4}$-$10^{13.6}$ & 1317 & $1.20_{-0.06}^{+0.06}$ & $0.64_{-0.02}^{+0.02}$ & $0.9969_{-0.0008}^{+0.0008}$ \\[-1em]  \\ \cline{2-6}
    \\[-1em]  & $10^{13.6}$-$10^{13.8}$ & 1290 & $1.18_{-0.07}^{+0.07}$ & $0.90_{-0.02}^{+0.02}$ & $0.9987_{-0.0010}^{+0.0010}$ \\[-1em]  \\ \cline{2-6}
    \\[-1em]  & $10^{13.8}$-$10^{14.0}$ & 1285 & $1.29_{-0.08}^{+0.08}$ & $1.19_{-0.03}^{+0.03}$ & $0.9987_{-0.0013}^{+0.0012}$ \\ \\[-1em] \hline \hline
    
    \\[-1em]
    \multicolumn{6}{|c|}{Simplified model}\\ \hline \hline
    \\[-1em] 
    \multirow{6}{*}{\begin{tabular}[c]{@{}c@{}}Protohalos\end{tabular}} & $10^{13.0}$-$10^{13.2}$ & 1256 & - & $0.39_{-0.02}^{+0.02}$ & $1.0011_{-0.0070}^{+0.0071}$ \\[-1em]  \\ \cline{2-6}
    \\[-1em]  & $10^{13.2}$-$10^{13.4}$ & 1289 & - & $0.57_{-0.02}^{+0.02}$ & $1.0036_{-0.0051}^{+0.0050}$ \\[-1em]  \\ \cline{2-6}
    \\[-1em]  & $10^{13.4}$-$10^{13.6}$ & 1268 & - & $0.81_{-0.02}^{+0.02}$ & $0.9916_{-0.0046}^{+0.0046}$ \\[-1em]  \\ \cline{2-6}
    \\[-1em]  & $10^{13.6}$-$10^{13.8}$ & 1284 & - & $1.09_{-0.03}^{+0.03}$ & $1.0000_{-0.0046}^{+0.0045}$ \\[-1em]  \\ \cline{2-6}
    \\[-1em]  & $10^{13.8}$-$10^{14.0}$ & 1283 & - & $1.40_{-0.04}^{+0.04}$ & $0.9992_{-0.0051}^{+0.0051}$ \\ \hline \hline
    \\[-1em] \multirow{5}{*}{\begin{tabular}[c]{@{}c@{}} \\ Halos\end{tabular}} & $10^{13.0}$-$10^{13.2}$ & 1360 & - & $0.41_{-0.02}^{+0.02}$ & $1.0014_{-0.0024}^{+0.0024}$ \\[-1em]  \\ \cline{2-6}
    \\[-1em]  & $10^{13.2}$-$10^{13.4}$ & 1313 & - & $0.56_{-0.02}^{+0.02}$ & $0.9998_{-0.0027}^{+0.0027}$ \\[-1em]  \\ \cline{2-6}
    \\[-1em]  & $10^{13.4}$-$10^{13.6}$ & 1290 & - & $0.76_{-0.02}^{+0.02}$ & $0.9985_{-0.0030}^{+0.0030}$ \\[-1em]  \\ \cline{2-6}
    \\[-1em]  & $10^{13.6}$-$10^{13.8}$ & 1239 & - & $1.02_{-0.03}^{+0.03}$ & $1.0016_{-0.0034}^{+0.0034}$ \\[-1em]  \\ \cline{2-6}
    \\[-1em]  & $10^{13.8}$-$10^{14.0}$ & 1250 & - & $1.23_{-0.04}^{+0.04}$ & $1.0018_{-0.0041}^{+0.0041}$ \\  \hline \hline
    \\[-1em] \multirow{5}{*}{\begin{tabular}[c]{@{}c@{}}\\Protohalos \\+ $\delta_{\rm{m}}$ \end{tabular}} & $10^{13.0}$-$10^{13.2}$ & 1601 & - & $0.45_{-0.01}^{+0.01}$ & $1.0005_{-0.0006}^{+0.0006}$ \\ \cline{2-6}
    \\[-1em]  & $10^{13.2}$-$10^{13.4}$ & 1855 & - & $0.67_{-0.01}^{+0.01}$ & $1.0011_{-0.0006}^{+0.0006}$ \\[-1em]  \\ \cline{2-6}
    \\[-1em]  & $10^{13.4}$-$10^{13.6}$ & 1655 & - & $0.88_{-0.01}^{+0.01}$ & $0.9976_{-0.0008}^{+0.0008}$ \\[-1em]  \\ \cline{2-6}
    \\[-1em]  & $10^{13.6}$-$10^{13.8}$ & 1565 & - & $1.17_{-0.01}^{+0.01}$ & $0.9999_{-0.0009}^{+0.0010}$ \\[-1em]  \\ \cline{2-6}
    \\[-1em]  & $10^{13.8}$-$10^{14.0}$ & 1531 & - & $1.52_{-0.02}^{+0.02}$ & $1.0002_{-0.0012}^{+0.0012}$  \\ \hline \hline
            \end{tabular}
            }
\end{table}

We obtain the best fit values of $b_{10}$, $b_{01}$ and $\alpha$ with Markov Chain Monte Carlo (MCMC), implemented in {\texttt{emcee}} \citep{Foreman-Mackey2013}. The full model has 1172 degrees of freedom: 78 free parameters (including 75 nuisance parameters), 50 radial bins and 25 simulation boxes. Table \ref{tab:merged fit results} summarizes our fit results. Our $\chi^2$ values are consistent with the expectation given the number of degrees of freedom. The protohalo+ $\delta_{\rm m}^W$field produces higher $\chi^2$ values, since we don't explicitly model the smoothing, but just allow the bias parameters to absorb its effect.

Fig.~\ref{fig:alpha best fit} shows the fit results for $\alpha$. 
Protohalos + $\delta_{\rm m}$ offer the best precision in $\alpha$: the errors on the $\alpha$ measured in the combined field are 3.5 times smaller than in halos -- and 4-8 times smaller than in protohalos alone.
$\alpha$ is consistent with an unbiased ruler within 1$\sigma$ in halos. For protohalos, protohalos + $\delta_{\rm{m}}$, and protohalos + $\delta_{\rm{m}}^W$, $\alpha$ is consistent with an unbiased ruler within 2$\sigma$ across all but the $10^{13.4}-10^{13.6}\ M_\odot/h$ mass bin. 
Errors on $\alpha$ increase monotonically with mass in halos and the combined field; shot noise is a likely driver of this increase. $\alpha$ errors show no mass dependence in protohalos alone, which we can attribute to the protohalo 2PCF's low signal-to-noise, as evidenced by the solid line in Fig.~\ref{fig:merged results} -- and predicted in Section~\ref{sec:scale-dep bias}.
When  $\delta_{\rm{m}}^W$ is smoothed by $R_\delta=1$ Mpc/$h$ (shown in Table \ref{tab:merged fit results}), $\alpha$ errors increase by 23\%, averaged over mass bins. At $R_\delta=2$ Mpc/$h$, $\alpha$ errors increase by a further 5\%. While smoothing $\delta_{\rm m}$ impacts the errors, the median $\alpha$ values are not affected.

Fig.~\ref{fig:merged results} compares the best-fit model to the mean {\sc{AbacusSummit}} 2PCF. The halo and protohalo + $\delta_{\rm m}$ columns have the same vertical axis ranges, which validates our Sec.~\ref{sec:scale-dep bias} prediction that the addition of $\delta_{\rm{m}}$ to protohalos boosts the 2PCF to match the halo 2PCF's amplitude, while creating a sharper peak. $\xi_{\rm{Lin}}$ highlights the protohalo 2PCF's low amplitude, as well as the difference in the shapes of the halo and combined-field 2PCFs.

Fig.~\ref{fig:bias values summary} shows the fit results for $b_{10}$ and $b_{01}$. $b_{10}$ values are consistent across all tracers. Our Eulerian-space measurements of $b_{01}$  fall far above what is expected from peaks theory (Sec.~\ref{sec: comparison to theory}) and decrease with mass, contrary to expectation.
Thus, the $b_{01}$ values in Table~\ref{tab:merged fit results} are not a faithful representation of scale-dependent bias. Since there is a substantial degeneracy between the $k^2$ bias term and the Gaussian smearing term in Eq.~\ref{eqn:halo overdensity}, we ran a diagnostic Eulerian fit with the smearing scale in Eq.~\ref{eqn:halo overdensity} adjusted by 25\%. A 25\% decrease in $R_{\rm smear}$ brought the best-fit $b_{01}$ (and $b_{10}$) values very close to those measured with protohalos, with no significant change in $\alpha$ or $\chi^2$. Increasing $R_{\rm smear}$ by 25\% had the opposite effect on $b_{01}$, with a notable increase in $\chi^2$.

The protohalo + $\delta_{\rm{m}}^W$ field produced lower $b_{01}$ than protohalos alone or the unsmoothed combined field. This can be explained by the mismatch in the shapes of the $\delta_{\rm{m}}^W$ 2PCF and $\xi_{\rm{Lin}}$ due to the smoothing. An analysis that relies on a smoothed matter field should adjust their model 2PCF accordingly. However, our motivation for considering the smoothed mass density field was as a crude model of the reconstructed matter density field \cite{Nikakhtar2024}. Given that our scope here was to consider the idealized protohalo case, we do not explore adjusting the template.

\subsection{Fitting a model without a scale-dependent term}\label{sec:fitting without b01}

The core shape of the standard BAO template used by large-scale surveys is driven by $\xi_{\rm{Lin}}(r)$ \cite[e.g.][]{Bautista2020}. To put Sec.~\ref{sec:full model fit}'s results in the context of a $\xi_{\rm{Lin}}$-driven template, we compare the performance of the full model developed in Sec.~\ref{sec:scale-dep bias} to a simplified model, where $b_1(k)$ does not include a $k^2$ term. This changes the protohalo overdensity relation from Eq.~\ref{eqn: ph overdensity} to
\begin{equation}\label{eqn:no b01 ph overdensity}
    \delta_{\textrm{ph}}(\vec{k})=b_{10}(\vec{k})\Tilde{W}(k R_f)\delta_{\textrm{m}}(\vec{k}).
\end{equation}

The halo overdensity relation (Eq.~\ref{eqn:no b01 halo overdensity}) becomes
\begin{equation}\label{eqn:no b01 halo overdensity}
    \delta_{\textrm{h}}(\vec{k})=(b_{10} +1) \Tilde{W}(kR_{\rm smear})\delta_{\textrm{m}}(\vec{k}),
\end{equation}
and for the protohalo$+\delta_{\rm{m}}$ field, Eq.~\ref{eqn: combined overdensity} simplifies to
\begin{equation}\label{eqn:no b01 combined overdensity}
\delta_{\textrm{comb}}(\vec{k}) = \left[ b_{10}\Tilde{W}(kR_f) + 1\right] \delta_{\textrm{m}}(\vec{k}).
\end{equation}

We repeat the procedure introduced in Sec.~\ref{sec:full model fit} with $\xi_{\rm t}$ built on the expressions above. Removing $b_{01}$ from the model increases the degrees of freedom to 1173.  Visual inspection of Fig.~\ref{fig:merged results} reveals that the simplified model does not capture the shape of the data 2PCF as well as the full model for any mass bin or tracer type. $\chi^2$ values in Table~\ref{tab:merged fit results} reflect this decreased quality of fit. The residuals show the simplified model's increased departures from the data, notably around the BAO peak. Although most apparent in the combined field, this feature is present in all tracers and results from the template not accurately capturing the shape of the BAO feature. The full model's residuals, on the other hand, show no pattern apart from statistical noise. This supports the idea that a $k^2$ bias term makes a valuable contribution to a model that captures the shape of the biased 2PCF. 

Fig.~\ref{fig:alpha best fit} shows that the $\alpha$ values measured with the  simplified model fall well within 1$\sigma$ of the full model. This supports the assumption behind the BAO analyses that only include a constant bias term in their models: although the shape of the simplified model deviates from data, this difference does not affect the best-fit value of the BAO scale. However, the simplified model's errors on $\alpha$ are larger for all of the cases we tested. Averaged across mass bins, the improvement in $\alpha$ errors offered by the full model is 47\% in protohalos alone, 15\% in halos, and 14\% in the combined protohalo + $\delta_{\rm{m}}$ field. 

Our findings about the importance of scale-dependent bias impact other cosmological distance scale estimators, including the scale $r_0$ at which $\xi_{\rm Lin}$ changes sign (i.e. crosses 0), and the Linear Point $r_{\rm{LP}}$ -- the midpoint between the BAO peak and dip scales. While evidence supports that $r_{\rm{LP}}$ is quite robust to $k^2$-bias \cite{Anselmi2016}, scale-dependent bias can strongly affect $r_0$. We discuss this further in Appendix~\ref{appx:r0}.

\subsection{Comparison to theoretical bias value predictions}\label{sec: comparison to theory}

Bias parameters satisfy a hierarchy of consistency relations that check the salience of best-fit values \cite[e.g.][]{Musso2012, Paranjape2013}. These relations are derived from the constraints that define a protohalo in a given model of halo formation \cite[for a straightforward method, see][]{Chan2017}. The constraint that a peak's height must exceed a threshold overdensity $\delta_c$ to become the site of a protohalo leads to the simple relation
\begin{equation}\label{eqn:consistency relation}
    b_{10}+b_{01}=\frac{\delta_c}{\sigma_0^2(R_f)},
\end{equation}
where we approximate $\delta_c=1.686$, the overdensity threshold for the spherical collapse of a halo. Since protohalos are not spherical in reality, this value serves as an approximate check on the self-consistency of our bias values. The right panel in Fig.~\ref{fig:bias values summary} compares the sum of best-fit values of $b_{10}$ and $b_{01}$ to the value predicted by the consistency relation for $R_f$ values that correspond to the median protohalo mass for each bin in our work.

For massive halos, the sum of the Lagrangian bias factors agrees with Eq.~\ref{eqn:consistency relation}, with $\delta_c = 1.686$.  However, agreement at smaller masses would require $\delta_c$ to increase.  The required increase is in qualitative agreement with direct measurements of the mass dependence of the average overdensity within protohalo patches \cite{Sheth2001, Sheth2013, Chan2017}. Eulerian-space values fall very far from the consistency relation, given the inflated values of $b_{01}$, as discussed in Sec.~\ref{sec:full model fit}.

\subsection{Comparison to Fourier space}\label{sec:fourier}

\begin{figure}
    \centering
    \includegraphics[width=\columnwidth]{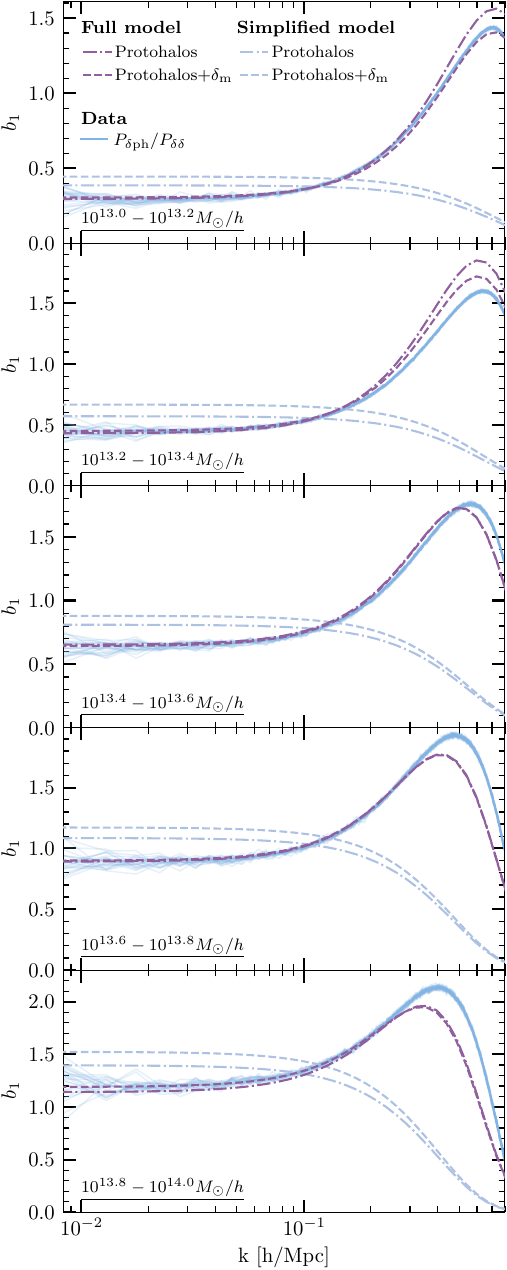}
    \caption{$b_1(k)$: measured in {\sc AbacusSummit} per Eq.~\ref{eq:b1} (solid blue) and the full (purple) and simplified (blue) models, with bias values from Table~\ref{tab:merged fit results}. Protohalo + $\delta_{\rm m}$ curves  are offset by $-1$ for a direct comparison with the protohalo $b_1(k)$.}
    \label{fig:b1_k_space}
\end{figure}

To compare the bias measured in protohalos and  protohalos + $\delta_{\rm m}$ more directly than is possible with the 2PCF, we turn to Fourier space. Linear bias $b_1(k)$ is simple to estimate in Fourier space from the ratio of the protohalo-matter cross-spectrum to the matter auto-spectrum:
\begin{equation}\label{eq:b1}
b_1(k)=\frac{P_{\delta \, \mathrm{ph}} (k)}{P_{\delta \delta} (k)},
\end{equation}
where $\delta$ is the Lagrangian overdensity field scaled to $z^E$ using linear theory. Fig.~\ref{fig:b1_k_space} shows Eq.~\ref{eq:b1} measured directly in 25 {\sc{AbacusSummit}} boxes as solid blue lines, which converge at high values of $k$. We compare the {\sc{AbacusSummit}} $b_1(k)$ to our models of $b_1(k)$ (Table~\ref{tab:b1}), using best-fit $b_{10}$ and $b_{01}$ values measured in protohalos and protohalos + $\delta_{\rm m}$ (Table~\ref{tab:merged fit results}). Note that we have subtracted 1 from the protohalo + $\delta_{\rm m}$ curves to enable a direct comparison to the protohalo-only curves.

The full model applied to the protohalo + $\delta_{\rm{m}}$ field offers the closest match to the data $b_1(k)$, with protohalos alone providing a match that is almost as close. The simplified model misses the data $b_{1}(k)$ for both the combined field and protohalos alone. Although the simplified model only includes the constant bias factor $b_{10}$, it turns over because of the window function.

\section{Conclusions}\label{sec: conclusions}

This work revisits the measurement of the BAO scale, focusing on two areas: protohalos and scale-dependent bias. A method rooted in optimal transport theory has given rise to the possibility of reconstructing the positions of protohalos along with the underlying matter density field \cite{Nikakhtar2024}. We fit a model to the correlation function of protohalos -- and introduce a new approach to this task by combining protohalos with the matter overdensity field $\delta_{\rm{m}}$. We compare the robustness of the BAO scale parameter $\alpha$ in protohalo and combined-field data to Eulerian-space measurements made with halos. Additionally, we have examined the performance of a template 2PCF built on a tracer bias relation that includes scale dependence -- which can be motivated by peaks theory or a general bias expansion.

This work has two main implications for future efforts to measure the BAO scale:
\begin{enumerate}
    \item The results here strongly motivate linearly combining the reconstructed halos with the full divergence-of-displacements field for BAO scale analyses. 
    \item We recommend incorporating scale-dependent bias into the template model.
\end{enumerate}

The precision of the BAO distance scale estimate is highest with the linear combination of the protohalo and matter fields, which offers a factor of 3.5 improvement over Eulerian-space measurements and a factor of 4-8 improvement over the estimate made with protohalos alone. Smoothing the $\delta_{\rm{m}}$ in the combined field by 1 Mpc/$h$ increases the errors by 23\%. In halos, $\alpha$ is consistent with an unbiased ruler within 1$\sigma$. In protohalos alone and the combined field, $\alpha$ is consistent with an unbiased ruler within 2$\sigma$ in four of the five mass bins we consider.

A model that includes $b_{01}$ outperforms a model without scale-dependent bias at capturing the shape of the 2PCF of biased tracers. Additionally, scale-dependent bias $b_{01}$ decreases the errors in $\alpha$. Allowing for scale dependence does not appear to introduce a shift in the BAO feature. Lagrangian-space data -- protohalos or a linear combination of protohalos and matter overdensity -- provides a sound estimate of the scale-dependent bias value, $b_{01}$. 

The match between the best-fit bias values and the values predicted theoretically (Section~\ref{sec: comparison to theory}) indicates that a bias model which includes a $k^2$ term is consistent with a simple structure formation picture such as the one proposed by extended peaks theory.

The presence of $k^2$-bias can profoundly impact other estimators of the cosmological distance scale.  In particular, it can significantly modify the zero-crossing scale, especially for the reconstructed halo field (Appendix~\ref{appx:r0}). Scale-dependent bias is less evident in the evolved field (Eq.~\ref{eqn:b01E}), and the zero-crossing for evolved tracers is therefore less biased.

Our results show that a scale-dependent bias term in $b_1(k)$ impacts the clustering signal, even in Eulerian space. However, this analysis works with narrow bins of mass. Whether or not the scale-dependent bias term's effect washes out when averaging over a wide range of masses is a question for a future work. Intuitively, one may expect that averaging over masses may have a relatively benign impact in Eulerian space. We base this supposition on the small range of $b(k)$ spanned by realistic Eulerian samples: the clustering amplitude of $10^{13.8}-10^{14.0}\ M_\odot /h$ halos is only about a factor of two and a half higher than that of $10^{13.0}-10^{13.2}\ M_\odot /h$ halos (middle column in Fig.~\ref{fig:merged results}). Most of the bias model's effect on the shape of the Eulerian 2PCF should therefore be attributed to the mass dependence of the smoothing scale $R_f$ and smearing scale $R_{\rm smear}$ (Table~\ref{tab:mass bin properties}) -- rather than the bias factors $b_{10}$ or $b_{01}$. However, at low redshifts, $R_f^2 \ll R_{\rm smear}^2$ -- and $R_{\rm smear}$ has only a weak mass dependence. Lagrangian space is more complex: the clustering signal can vary by much larger factors (the top and bottom protohalo panels in Fig.~\ref{fig:merged results} differ by a factor of 15), and because there has been no smearing, the mass dependence of $R_f$ is relevant.  We hope that our results motivate studies that efficiently parameterize the effect that mass-averaging would have on $b(k)$.

Expanding the protohalo + $\delta_{\rm{m}}$ formalism to measuring the BAO scale in observational data is a clear next step to follow from our work. This will necessitate reconstructing protohalo locations from observed galaxy-hosting halos. The possibility of doing so with a method built on optimal transport theory may pave a promising avenue for calibrating the BAO scale in galaxy survey data \cite{Nikakhtar2021}. Part of the expansion to observations should investigate the effect of noise on the $\delta_{\rm{m}}$ in the combined field, in addition to the smoothing that we have introduced here. 

The shape of the correlation function encodes a plethora of cosmological information that goes beyond the scale of the BAO feature -- the focus of our paper. For BAO scale analyses, ignoring the scale dependence of bias has the arguably benign effect of inflating the error bars -- without shifting the location of the BAO peak. However, Table~\ref{tab:merged fit results} and Fig.~\ref{fig:b1_k_space} show that omitting scale-dependent bias returns incorrect values of $b_{10}$, the scale-{\em in}dependent bias factor (compare Full and Simplified values of $b_{10}$ for halos).  This will compromise estimates of $A$, the amplitude of the matter fluctuation field.  Although allowing for nonzero $b_{01}$ returns an unbiased estimate of $b_{10}$, we hope that future work revisits $b_{01}$ from Eulerian-space fits to address the inflated fit values found here, as this will give more confidence in the consequent constraints on $A$.

\section*{Acknowledgements}
SG wishes to thank the ICTP, Trieste, for its hospitality during the 2022 Summer School on Cosmology, where some of the coauthor discussions took place. SG was supported in part by the National Science Foundation Graduate Research Fellowship Program under grant No. DGF-1752134. SG and NP were supported in part by DOE DE-SC0017660. FN gratefully acknowledges support from the Yale Center for Astronomy and Astrophysics Prize Postdoctoral Fellowship. RKS is grateful to the EAIFR, Kigali, and the IFPU and ICTP, Trieste for hospitality in 2024 when this work was completed.  

\appendix

\section{Covariance matrix}\label{sec:covariance matrix}
 
We obtain the smoothed binned covariance matrix $\underline{\underline{C}}$ following the formalism in \cite[e.g.][]{Smith2007, Sanchez2008, Smith2009,Xu2010, Grieb2016,  Lippich2019, Parimbelli2021}. A matrix element corresponding to radii $r_i$ and $r_j$ is given by
\begin{equation}\label{eqn:cov mx}
{C}_{i j}=\frac{2}{L_{\rm{box}}^3} \int_{\boldsymbol{k}}  \bar{j}_0\left(k r_i\right) \bar{j}_0\left(k r_j\right)(P(k) + \mathcal{N})^2,
\end{equation}
where $P(k)$ is the mean of the simulation power spectra for a given tracer and mass bin. $L_{\rm{box}}$ is the simulation box size. We fit our model to each realization's 2PCF individually, so Eq.~\ref{eqn:cov mx} is not normalized by sample size. Eq. \ref{eqn:cov mx} is the standard analytic Gaussian covariance matrix for the 2PCF of binned data. %Without binning, the standard matrix includes the zeroth-order spherical Bessel function of the first kind, $j_0(k r)$. 
To account for the binning, we average $j_0(k r)$ over of a single radial bin to get $\bar{j_0}(k r)$, a band-averaged spherical Bessel function for a bin centered on $r$ with edges $r_1$ and $r_2$, and $\Delta r = r_2-r_1$:
\begin{equation}
\bar{j}_0(k r)=\frac{r_2^2 j_1(k r_2)-r_1^2 j_1(k r_1)}{r^2 k \Delta r},
\end{equation}
where $j_1(kr)$ is the first-order spherical Bessel function of the first kind. $\mathcal{N}$, the shot noise, comprises two terms:
\begin{equation}
    \mathcal{N} =\bar{n}_h^{-1}+\epsilon.
\end{equation}
The linear shot noise, assumed to be Poisson,  equals the inverse of the (proto)halo number density $\bar{n}_h^{-1}$\citep{Cohn2006,Smith2008} (see Table~\ref{tab:number densities} for the mean {\sc AbacusSummit} values). $\epsilon$ is the nonlinear shot noise that we can attribute to nonlinear structure growth at small scales. We expect $\epsilon \ll \bar{n}_h^{-1}$, which turns out to be the case after the optimization below, starting with the likelihood of the dataset, $\mathcal{L}$ \citep{Xu2012}:
\begin{equation}
    \mathcal{L} = \prod_{i=0}^{N_{\rm{r}}} \left[  (2 \pi)^{q} \det \underline{\underline{C}}\ \rm{e}^{\chi^2_i} \right]^{-1/2}.
\end{equation}
We minimize $\mathcal{L}$'s negative log,
\begin{equation}
L=-2 \ln \mathcal{L}=q\, N_{\rm{r}}  \ln (2 \pi)+N_{\rm{r}}  \ln (\det\ \underline{\underline{C}})+\sum_{i=0}^{N_{\rm{r}}}  \chi_i^2,
\end{equation}
where $q$ is the number of points to fit, $N_{\rm{r}}$ is the number of boxes and  $\chi_i^2 = \vec{\Delta \xi_i}  \cdot \underline{\underline{C}}^{-1} \cdot \vec{\Delta \xi}^{\rm{T}}_i$, with $\vec{\Delta \xi_i} \equiv \vec{\xi}_{i}-\langle \vec{\xi} \rangle.$

\begin{table}
 \caption{Mean halo number densities in {\sc AbacusSummit}.}
\label{tab:number densities}
\begin{tabular}{ @{\hskip 0.15in}c@{\hskip 0.25in}c@{\hskip 0.15in} }
\hline  \hline \\[-1.04em]
Mass bin $[M_\odot/h]$ &
$\bar{n}_h/\left[ \rm{Mpc}/h \right]^{-3}$\\ \hline \\[-1em] $10^{13.0}$-$10^{13.2}$ &
    $1.670 \times 10^{-4}$ \\[-1em]  \\ 
    \\[-1em] $10^{13.2}$-$10^{13.4}$ &
    $1.046 \times 10^{-4}$ \\[-1em]  \\ 
    \\[-1em] $10^{13.4}$-$10^{13.6}$ &
    $6.396 \times 10^{-5}$ \\[-1em]  \\ 
    \\[-1em] $10^{13.6}$-$10^{13.8}$ &
    $3.788 \times 10^{-5}$ \\[-1em]  \\ 
    \\[-1em] $10^{13.8}$-$10^{14.0}$ &
    $2.141 \times 10^{-5}$ \\[-1em]  \\ \hline \hline
\end{tabular}
\end{table}

\begin{figure*}
    \centering
    \includegraphics[width=\linewidth]{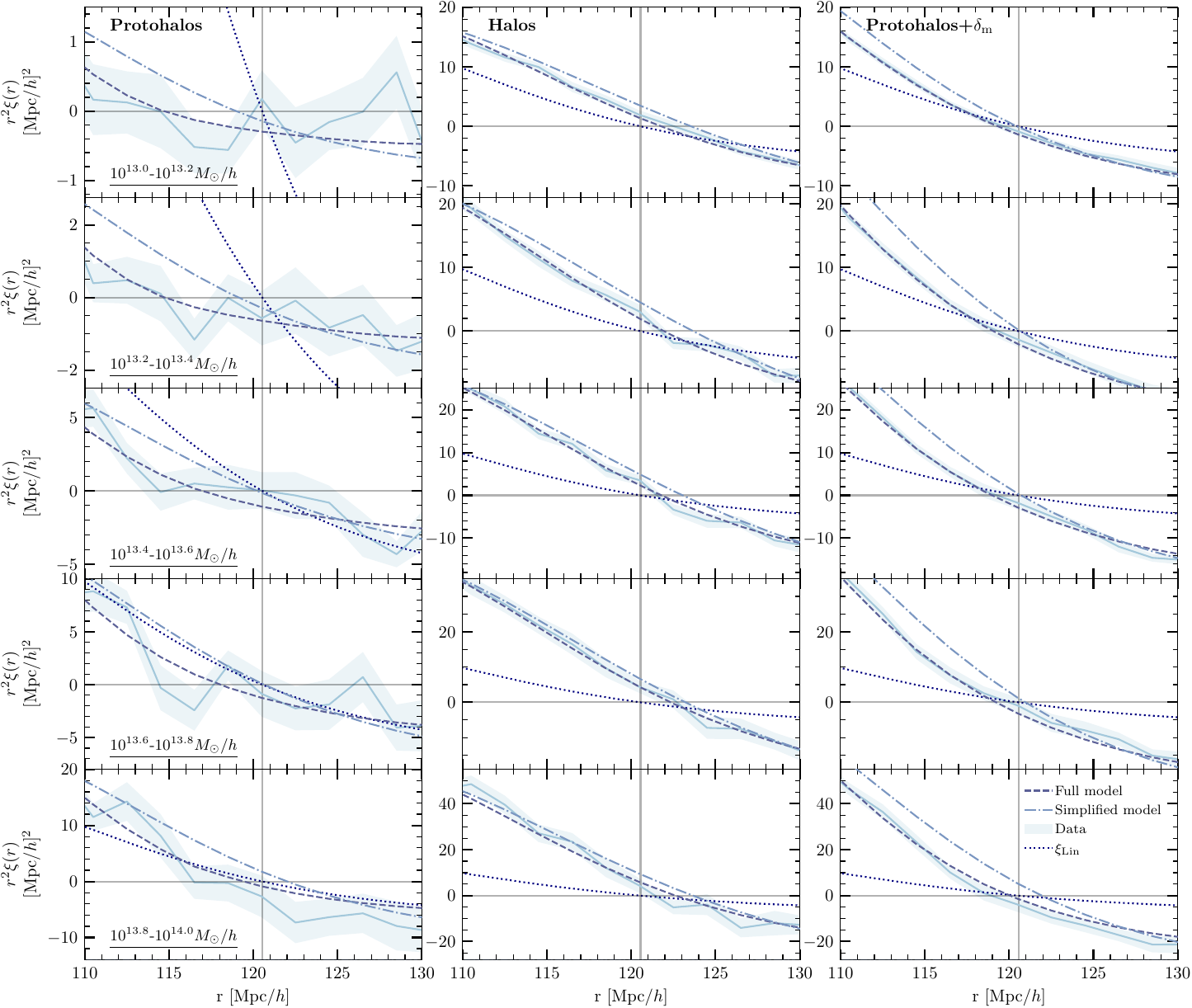}
\caption{Same as (odd rows in) Fig.~\ref{fig:merged results}, but now highlighting the scale $r_0$ where $\xi_{\rm{Lin}}$ (dotted curve, same in each panel) crosses zero (cross-hair, same in each panel).  For protohalos, the zero-crossing occurs on smaller scales, and is quite mass dependent (left panels); for Eulerian halos, the scale is less mass-dependent, but consistently slightly larger than $r_0$ (middle panels).  }
\label{fig:r0}
\end{figure*}

To speed up the minimization of $L$, we pre-compute the matrix $\underline{\underline{\kappa_l}}$, where each element equals
\begin{equation}
\kappa_{l, ij}= \frac{2}{L_{\rm{box}}^3} \int_{\boldsymbol{k}} \bar{j}_0\left(k r_i\right) \bar{j}_0\left(k r_j\right)(P(k) + \bar{n}_h^{-1})^l,
\end{equation}
which lets us write $\underline{\underline{C}}$ as the quadratic expression 
\begin{equation}
\underline{\underline{C}}= \underline{\underline{\kappa_2}} + 2 \epsilon \underline{\underline{\kappa_1}} + \epsilon ^ 2 \underline{\underline{\kappa_0}}.
\end{equation}
A given covariance matrix's value of $\epsilon$ depends on the mass, redshift and tracer type. The optimal values of $\epsilon$ fall within 1-3\% of $\bar{n}_h^{-1}$, as expected\footnote{In this work's narrow bins of mass, $\bar{n}_h^{-1}$ is only 1-2\% lower than mass-weighted shot noise, $\bar{n}_h^{-1}\left\langle m^2\right\rangle /\langle m\rangle^2$. However, optimal $\epsilon$ values do not equal this difference.}. 

 \section{Nuisance parameters}\label{sec:nuisance parameters}
We obtain the nuisance parameters in Eq.~\ref{eqn:standard template} analytically via marginalization, which has been shown to preserve information about non-nuisance parameters \cite{Taylor2010}. We solve for the column vector $\vec{a}$, where each element is a nuisance parameter, which maximizes the model's $\chi^2$:
\begin{align}\label{eqn:max chisq}
    \frac{\partial \chi^2}{\partial \vec{a}}&=\frac{\partial}{\partial \vec{a}} \left[\vec{\Delta \xi}^T \cdot \underline{\underline{C}}^{-1} \cdot \vec{\Delta \xi} \right] =0,
\end{align}
where $\vec{\Delta \xi}$ is a column vector with the difference between a given simulation box's 2PCF and the template (Eq.~\ref{eqn:standard template}): 
 \begin{equation}
     \vec{\Delta \xi}  \equiv \vec{\xi}_{i}-\vec{\xi}_{\rm{t}} (b_1, \alpha)-\vec{a}^{\ T} \underline{\underline{\mathcal{R}}},
 \end{equation}
with elements corresponding to $m$ radial bins. $\underline{\underline{\mathcal{R}}}$ is an $n \times m$ matrix with rows corresponding to nuisance functions:
\begin{equation}
    \underline{\underline{\mathcal{R}}} = \begin{bmatrix}
1 & 1 & \cdots & 1\\
r_1^{-1} & r_2^{-1} & \cdots & r_m^{-1}\\
\vdots & \vdots & \ddots & \vdots\\
r_1^{-(n-1)} & r_2^{-(n-1)} & \cdots & r_m^{-(n-1)}
\end{bmatrix}.
\end{equation}
We rearrange Eq.~\ref{eqn:max chisq} to get $\vec{a}$:
\begin{equation}\label{eqn: nuisance vector}
    \vec{a} = \left(\underline{\underline{\mathcal{R}}} \ \underline{\underline{C}}^{-1} \underline{\underline{\mathcal{R}}}^{T} \right) ^{-1} \underline{\underline{\mathcal{R}}} \ \underline{\underline{C}}^{-1} \cdot \left(\vec{\xi}_{i}-\vec{\xi}_{\rm{model}} (b_1, \alpha)\right).
\end{equation}
Pre-computing the component of Eq. \ref{eqn: nuisance vector} with $\underline{\underline{\mathcal{R}}}$ and $\underline{\underline{C}}^{-1}$  simplifies the linear algebra operations necessary to solve for 75 nuisance parameters at each likelihood evaluation.

\section{The 2PCF zero-crossing point}\label{appx:r0}
The main text focused on the standard methodology for estimating the BAO distance scale, in which a cosmological model-based template is fitted to the measured pair correlation function.  The zero-crossing of the pair correlation function is potentially an interesting scale, because zero-crossing can be estimated {\em without} fitting a cosmological-model based template, and any scale-independent bias will leave this scale unchanged \cite[e.g.][]{Klypin1994, Prada2011}.  However, our demonstration that scale-dependent bias has a significant impact on the protohalo distribution, and, to a lesser extent, the evolved halo field as well, suggests that the zero-crossing may not be as pristine a ruler as one might have hoped.  

To address this, Fig.~\ref{fig:r0} shows the scales around which the measurements cross zero for protohalos, halos, protohalos + $\delta_{\rm m}$, and linear theory. The vertical line shows the zero-crossing scale $r_0$; this is the scale that is potentially a standard ruler.  However, for protohalos, the zero-crossing occurs on substantially smaller scales than this, by an amount that depends on halo mass.  Therefore, if one were to fit, e.g., polynomials, to the protohalo measurements, and estimate the zero-crossing from the fit, one would {\em not} recover $r_0$.  

The mismatch is a result of the scale-dependence of bias.  The dashed curve in each panel shows that the model that includes scale-dependent bias successfully recovers this (tracer-dependent) shift. Although the main text shows that a model with no scale-dependent bias produces a worse fit -- and Fig.~\ref{fig:r0} shows that it does {\em not} describe the zero-crossing of the measurements -- its zero-crossing  is substantially closer to linear theory.  This is approximately true for the protohalos + $\delta_{\rm m}$ field as well.  Hence, at least in the reconstructed field, one could fit this poorer model to the measurements and then estimate the zero-crossing from it, but doing so is clearly not optimal.  
In summary, at least in the reconstructed field of realistically biased tracers, the zero-crossing is not as model-independent a ruler as one might have hoped.  

The main text showed that scale-dependent bias is less of an issue for Eulerian halos.  The middle panels of Fig.~\ref{fig:r0} show that this is also true for the zero-crossing: compared to the protohalo panels, the dashed curves are closer to the linear $r_0$, and the dot-dashed curves are not far off.  In addition, the measurements cross zero on a scale that is about 1.5-3\% higher than the linear theory value in all cases, with the discrepancy increasing  with mass.  The slight increase in scale is due to the smearing of the BAO feature.  The top panel of Fig.~\ref{fig:subcomponent} shows the effect of smearing when scale-dependent bias is ignored:  larger smoothing shifts the zero-crossing to larger scales.

Our results suggest that if one wants an estimate of $r_0$ that is not tied to the shape of $\xi_{\rm Lin}$, then one should {\em not} work with the reconstructed biased tracers:  one should either work with the Eulerian tracer field (in which case one must account for a slightly overestimated $r_0$), or with the full reconstructed dark matter field.

\bibliographystyle{apsrev4-2}
\bibliography{gaines2024_protohalo_bib}

\end{document}